\documentclass{INTERSPEECH2024}

\interspeechcameraready

\usepackage{multicol}
\usepackage{multirow}
\usepackage{color, colortbl}
\usepackage{xcolor}
\colorlet{eng}{blue!10}
\colorlet{cmn}{teal!10}
\colorlet{multi}{yellow!10}
\colorlet{euro}{orange!10}
\usepackage{pifont}
\usepackage{hyperref}
\usepackage{graphicx}
\usepackage{subcaption}
\usepackage{tablefootnote}
\title{SingOMD: Singing Oriented Multi-resolution Discrete Representation Construction from Speech Models}

\makeatletter
\def\thanks#1{\protected@xdef\@thanks{\@thanks
        \protect\footnotetext{#1}}}
\makeatother

\name[affiliation={1}]{Yuxun}{Tang}
\name[affiliation={1}]{Yuning}{Wu}
\name[affiliation={2}]{Jiatong}{Shi}
\name[affiliation={1}]{Qin}{Jin*}\thanks{ *Corresponding Author.}
\address{ $^{1}$ Renmin University of China,     $^{2}$ Carnegie Mellon University
}
\email{tangyuxun@ruc.edu.cn}
\email{\{tangyuxun, yuningwu, qjin\}@ruc.edu.cn, jiatongs@cs.cmu.edu}

\usepackage[
backend=biber,
style=ieee,
citestyle=numeric-comp,
maxbibnames=10,
maxcitenames=20,
doi=false,isbn=false,url=false,eprint=false
]{biblatex}

\defbibheading{bibliography}[\refname]{}

\DeclareSourcemap{
	\maps[datatype=bibtex, overwrite=true]{
		\map{
			\step[fieldsource=booktitle,
			match=\regexp{.*Interspeech.*},
			replace={Proc. Interspeech}]
			\step[fieldsource=journal,
			match=\regexp{.*INTERSPEECH.*},
			replace={Proc. Interspeech}]
			\step[fieldsource=booktitle,
			match=\regexp{.*ICASSP.*},
			replace={Proc. ICASSP}]
			\step[fieldsource=booktitle,
			match=\regexp{.*icassp_inpress.*},
			replace={Proc. ICASSP (in press)}]
			\step[fieldsource=booktitle,
			match=\regexp{.*Acoustics,.*Speech.*and.*Signal.*Processing.*},
			replace={Proc. ICASSP}]
			\step[fieldsource=booktitle,
			match=\regexp{.*International.*Conference.*on.*Learning.*Representations.*},
			replace={Proc. ICLR}]
			\step[fieldsource=booktitle,
			match=\regexp{.*International.*Conference.*on.*Computational.*Linguistics.*},
			replace={Proc. COLING}]
			\step[fieldsource=booktitle,
			match=\regexp{.*SIGdial.*Meeting.*on.*Discourse.*and.*Dialogue.*},
			replace={Proc. SIGDIAL}]
			\step[fieldsource=booktitle,
			match=\regexp{.*International.*Conference.*on.*Machine.*Learning.*},
			replace={Proc. ICML}]
			\step[fieldsource=booktitle,
			match=\regexp{.*North.*American.*Chapter.*of.*the.*Association.*for.*Computational.*Linguistics:.*Human.*Language.*Technologies.*},
			replace={Proc. NAACL}]
			\step[fieldsource=booktitle,
			match=\regexp{.*Empirical.*Methods.*in.*Natural.*Language.*Processing.*},
			replace={Proc. EMNLP}]
			\step[fieldsource=booktitle,
			match=\regexp{.*Association.*for.*Computational.*Linguistics.*},
			replace={Proc. ACL}]
			\step[fieldsource=booktitle,
			match=\regexp{.*Automatic.*Speech.*Recognition.*and.*Understanding.*},
			replace={Proc. ASRU}]
			\step[fieldsource=booktitle,
			match=\regexp{.*Spoken.*Language.*Technology.*},
			replace={Proc. SLT}]
			\step[fieldsource=booktitle,
			match=\regexp{.*Speech.*Synthesis.*Workshop.*},
			replace={Proc. SSW}]
			\step[fieldsource=booktitle,
			match=\regexp{.*workshop.*on.*speech.*synthesis.*},
			replace={Proc. SSW}]
			\step[fieldsource=booktitle,
			match=\regexp{.*Advances.*in.*neural.*information.*processing.*},
			replace={Proc. NeurIPS}]
			\step[fieldsource=booktitle,
			match=\regexp{.*Advances.*in.*Neural.*Information.*Processing.*},
			replace={Proc. NeurIPS}]
			\step[fieldsource=booktitle,
			match=\regexp{.*Workshop.*on.* Applications.* of.* Signal.*Processing.*to.*Audio.*and.*Acoustics.*},
			replace={Proc. WASPAA}]
			\step[fieldsource=publisher,
			match=\regexp{.+},
			replace={{}}]
			\step[fieldsource=month,
			match=\regexp{.+},
			replace={{}}]
			\step[fieldsource=location,
			match=\regexp{.+},
			replace={{}}]
			\step[fieldsource=address,
			match=\regexp{.+},
			replace={{}}]
			\step[fieldsource=organization,
			match=\regexp{.+},
			replace={{}}]
   \step[fieldsource=pages,
			match=\regexp{.+},
			replace={{}}]
		}
	}
}


\begin{document}

\maketitle
 
\begin{abstract}
Discrete representation has shown advantages in speech generation tasks, wherein discrete tokens are derived by discretizing hidden features from self-supervised learning (SSL) pre-trained models. However, the direct application of speech SSL models to singing generation encounters domain gaps between speech and singing. Furthermore, singing generation necessitates a more refined representation than typical speech. To address these challenges, we introduce SingOMD, a novel method to extract singing-oriented multi-resolution discrete representations from speech SSL models. Specifically, we first adapt the features from speech SSL through a resynthesis task and incorporate multi-resolution modules based on resampling to better serve singing generation. These adapted multi-resolution features are then discretized via clustering. Extensive experiments demonstrate the robustness, efficiency, and effectiveness of these representations in singing vocoders and singing voice synthesis.

\end{abstract}

\section{Introduction}
\label{sec: intro}

%%%% Para 1 %%%%%
Singing Voice Synthesis (SVS) has attracted considerable attention for its capability to produce high-fidelity vocal renditions from musical scores (e.g., lyrics, pitch and tempo).
A typical SVS framework is the cascaded system~\cite{seq2seqmodel, lu2020xiaoicesing, chen2020hifisinger, liu2022diffsinger}, where mel-spectrograms are initially generated in the acoustic model and subsequently employed by a vocoder to produce the singing waveform. 
Recently, in light of the advantages associated with discrete representations, such as reduced storage requirements, enhanced training efficiency, and potential compatibility with other modalities such as text, and coupled with related explorations in various speech-related tasks~\cite{lee-etal-2022-direct, yang2023universal, wang2023selm, huang2023makeavoice, chang2023exploring, barrault2023seamless, shi2023enhancing, wang2023neural, shi2023bridging, hayashi2020discretalk, shi2021discretization}, the concept of discrete SVS has begun to gain traction.
To further facilitate the exploration of discrete SVS, the Interspeech 2024 Challenge on speech processing using discrete units\footnote{\scriptsize{\url{}https://www.wavlab.org/activities/2024/Interspeech2024-Discrete-Speech-Unit-Challenge/}} has recently proposed the SVS track, aiming at utilizing discrete representations to construct the SVS system. Contestants are tasked with developing a cascaded discrete SVS system akin to those utilizing mel-spectrograms, including an acoustic model to convert the music score into discrete representation and a vocoder to transform the representation into waveform.

While the investigation into discrete representations within SVS is still in its nascent stages, notable advancements have been witnessed in speech generation domains, such as text-to-speech (TTS)~\cite{hayashi2020discretalk, huang2023makeavoice, borsos2023audiolm, yang2023universal, wang2023neural}, speech-to-speech translation~(S2ST)~\cite{lee-etal-2022-direct, wang2023speechtospeech, shi2023enhancing, barrault2023seamless}, speech enhancement (SE)~\cite{wang2023selm, shi2021discretization}. 
In these tasks, a prevalent method for obtaining discrete units involves conducting clustering over the intermediate features of speech self-supervised learning (SSL) pre-trained models~\cite{Mohamed_2022}.
Specifically, these pre-trained SSL models~\cite{hsu2021hubert, baevski2020wav2vec, chung2021w2vbert, chen2022wavlm, babu2021xls} typically operate as either speech encoders or frozen feature extractors, capturing latent features from the intermediate layers. Subsequently, clustering techniques such as K-means or Gumbel-Softmax are applied on these features to derive discrete representations, which are subsequently utilized in downstream tasks.

However, there are currently no SSL models specifically designed for singing-related tasks to extract representations suitable for singing, primarily due to constraints imposed by the scale of singing data~\cite{guo22e_interspeech, shi2021sequence}.
Leveraging speech SSL models could be a simple solution. However, considering the disparity between singing and speech domains, singing encompasses nuanced pitch variations, a broader spectrum of vocal frequencies, and longer durations, so singing synthesis requires richer and more expressive discrete representations. Therefore, direct application of discrete representations extracted from speech SSL models to singing-related tasks faces challenges of domain gap.
Moreover, inspired by the findings from Shi et al.~\cite{shi2024multiresolution, shi2023exploration} that a fixed resolution for speech signals is suboptimal, given that singing is more refined than speech, a single resolution is definitely not optimal for singing-related tasks. %This assertion stems from the limitation of a fixed field of view in adequately capturing information across different dimensions.

\begin{figure*}[t]
    \centering
    \includegraphics[width=\textwidth]{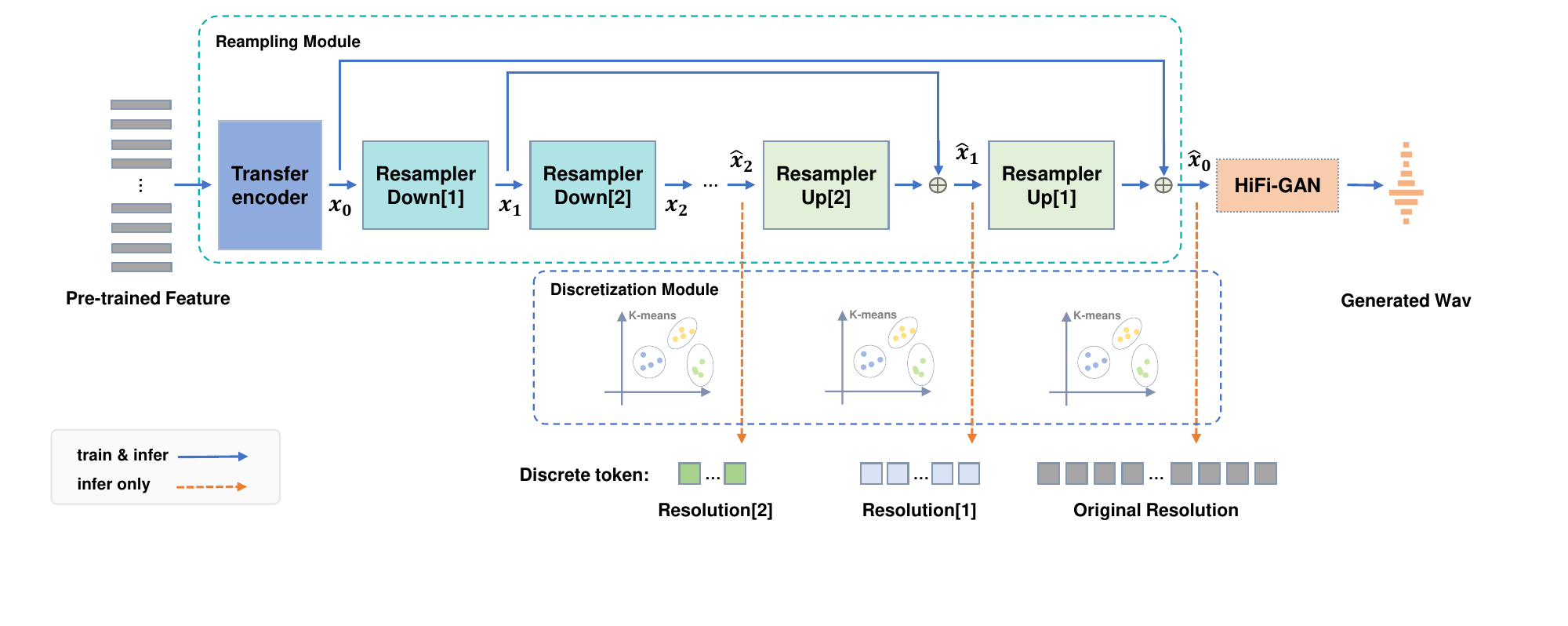}
    
    \vspace{-20pt}
    
    \caption{Illustration of the overall workflow of our proposed SingOMD.}
    \label{fig:singamd}
    
    % \vspace{-10pt}
\end{figure*}

%%%% Para 4 %%%%
To this end, we propose \textbf{SingOMD}, a novel method designed to extract \textbf{sing}ing \textbf{o}riented \textbf{m}ulti-resolution \textbf{d}iscrete representations by leveraging speech SSL models.
SingOMD trains continuous features extracted from raw singing audios using speech SSL models in a resynthesis task, thereby bridging the domain gap between the speech and singing. % to facilitate the transfer of features from the speech domain to the singing domain.
Moreover, to capture richer singing-specific features, we introduce a Unet-based resampling module~\cite{ronneberger2015unet, shi2024multiresolution} designed to incorporate multi-resolution features. Following the training of the resynthesis task, singing oriented multi-resolution continuous features are extracted from the intermediate layers of the resampling module, and subsequently clustered using K-means to obtain corresponding discrete representations.
Extensive experiments demonstrate that these singing oriented multi-resolution discrete representations exhibit high robustness in singing resynthesis. Furthermore, when integrated with a discrete singing voice acoustic model, our approach yields notable enhancements in both efficiency and effectiveness of singing voice synthesis.

The main contributions of this work include: (1) we propose a new method SingOMD to construct singing oriented multi-resolution discrete representations form speech SSL models using the resynthesis task without modifying model parameters or structures of speech SSL models; (2) our method achieves comparable results for singing voice synthesis with mel-spectrograms systems, and without using any auxiliary information, it outperforms the baseline which involves discrete representations from a larger model and requires pitch predictor. Our demo page can be accessed at \url{https://interspeech2024singomd.github.io}.

\section{Method}
%For the sake of clarity, the model trained using the SingAMD method will be referred to as SingAM. 

%\subsection{SingAMD}
Discrete representations have shown great potential in some speech related tasks but have not been well explored in SVS. 
%Because of the domain gap between speech and singing and the limitation of singing data scale, there is currently no  singing related SSL model, which poses a challenge to achieve good performance in singing-related task. 
As discussed in Section~\ref{sec: intro}, there is currently no singing-related SSL model, and leveraging speech SSL models faces challenges due to the domain gap between speech and singing.
Driven by issues above and inspiration from multi-resolution expolration in speech~\cite{shi2023exploration, shi2024multiresolution}, we propose a new method, SingOMD, to construct singing oriented multi-resolution discrete representations from speech SSL models.

%\subsubsection{Overview}
%We present SingAMD, a novel method to extract singing oriented discrete representations by leveraging speech SSL models in singing resynthesis task, as shown in 
Figure~\ref{fig:singamd} illustrates the overall workflow of our proposed SingOMD, which involves two major components: a resampling module to form multi-resolution, and a discretization module to produce multi-resolution discrete representations through clustering.
Initially, raw singing audio \vec{y} is fed into a frozen speech SSL model~($\mathrm{SSL}$) to obtain continuous speech features \Vec{s}, which is then fed into the resampling module~($\mathrm{Resampling}$) to produce features \vec{\hat{x}} in different resolution. Finally, the output features \vec{\hat{x}} from the resampling module are reconstructed to the waveform \vec{\hat{y}} using the $\mathrm{Vocoder}$. 
\begin{align}
 \vec{s} & = \mathrm{SSL}~(\vec{y})  \\
 \vec{\hat{x}} & = \mathrm{Resampling}~(\vec{s}) \\
 \vec{\hat{y}} & = \mathrm{Vocoder}~(\vec{\hat{x}}) 
\end{align}

Such singing audio resynthesis process can serve as an adapter to better fit the speech continuous features for singing. We then construct discrete representations by clustering on the continuous features \vec{\hat{x}}. We elaborate the working details of \mbox{SingOMD} below.
%To obtain discrete representations in multi-resolution, K-means clustering is performed over every resolution feature needed, which is extracted from each corresponding layer within the resampling module.

%\subsubsection{Architecture}

%When converting continuous features into discrete ones via clustering, an inevitable consequence is the loss of a substantial amount of information. 
%To mitigate this and maximize the information retained in the clustered features, SingAMD implements two primary strategies: firstly, enhancing the information content of the continuous features to the greatest extent feasible, and secondly, reducing the information loss during the clustering process as much as possible. These approaches are directly aimed at preserving the richness and detail of the original data, ensuring that the discrete representations are as informative and useful as possible.

\noindent \textbf{Speech Self-supervised Learning Model.} 
As the singing audio is fed into the Speech SSL model to obtain the continuous features, it is important to utilize as much of the available information as possible to augment the obtained features. In this context, following previous works~\cite{yang2021superb, shi2023mlsuperb}, we perform a weighted sum of the features extracted from all intermediate layers of the Speech SSL model. Let ${s_1, s_2, s_3, \ldots, s_n}$ represent these features from different hidden layers and $n$ is the number of layers. The obtained speech features $\vec{s}$ can be expressed as: $\vec{s} = \sum_{i=1}^{n}{w_i \cdot s_i}$, where $w_i$ denotes the trainable weight associated with the feature $s_i$. %and both of them are all trainable.

\noindent \textbf{Resampling Module.} 
%Regarding the latter point, to preserve information across different resolutions and ensure the richness of information, after obtaining the pre-trained speech features $\vec{s}$, we input them into a resampling module to generate intermediate representations at different resolutions. 
Inspired by findings from previous works \cite{shi2024multiresolution, shi2023exploration} that multi-resolution features are beneficial for speech processing tasks, we generate features at different resolutions based on the pre-trained speech features $\vec{s}$ in the resampling module, which 
%Within the resampling module, there are primarily 
involves a transfer encoder and downsampling-upsampling processing stages. 
The transfer encoder consists of a \texttt{Conv1d} layer with an equal number of input and output feature dimensions. 
Each resampler block in the downsampling-upsampling process is consistent with the sampling module described in \cite{shi2024multiresolution}, comprising a \texttt{Conv1d} for upsampling and a \texttt{ConvTranspose1d} for downsampling, cascaded in a residual manner. 
%In the specific process of resampling, initially, 
Specifically, the speech features $s$ first undergo a transfer encoder to adapt from the speech domain to the singing domain, producing singing features $x^{(0)}$ in the original resolution. 
These original singing features are then fed into the Unet-based upsampling-downsampling module. Firstly, they traverse through $t$ downsampling %modules $\text{down}[i]$ 
steps DOWN[i] sequentially to %downsample the singing features $x^{(i-1)}$, 
yield a sequence of downsampled features $x^{(0)}, x^{(1)}, \ldots, x^{(t)}$. %Subsequently, the upsampled feature $\hat{x^{(t)}} = x^{(t)}$ is inputted into the upsampling module $\text{up}[t]$, while introducing residuals to obtain upsampled information $\hat{x}^{(t-1)} = w_{res} \cdot (\hat{x}^{(t)} + x^{(t-1)})$. This process is symmetrically repeated to obtain upsampled features $\hat{x}^{(t)}, \hat{x}^{(t-1)}, \ldots, \hat{x}^{(0)}$. 
Subsequently, these features go through $t$ upsampling steps UP[i] that introduce residuals to obtain upsampled features  $\hat{x}^{(t-1)} = w_{\text{res}} \cdot (\hat{x}^{(t)} + x^{(t-1)})$, where $w_{\text{res}}$ is a hyperparameter. The downsampling and upsampling process is symmetrically repeated to obtain upsampled features $\hat{x}^{(t)}, \hat{x}^{(t-1)}, \ldots, \hat{x}^{(0)}$. 
The final features fed into the vocoder are $\vec{\hat{x}} = \hat{x}^{(0)}$.

%To better understand the process of upsampling and downsampling, let's provide further explanation. Firstly, 
Note that the scaling ratios of the upsampling and downsampling steps are symmetric. For instance, if $\text{DOWN}[1]$ downsamples by a factor of 2, then $\text{UP}[1]$ upsamples by a factor of 2, and this pattern continues. Additionally, the sampling ratios are determined by the adjacent sampling rates in the desired sampling rate sequence. For example, if the original audio resolution is 20ms and we aim to obtain features at [20ms, 40ms, 80ms], where 20ms represents the duration of each token (i.e., 50 features correspond to 1s of audio), then the downsampling ratios would be $[\frac{40}{20} = 2, \frac{80}{40} = 2]$. Furthermore, the upsampling ratios are symmetric to the downsampling ratios $[\frac{80}{40} = 2, \frac{40}{20} = 2]$.

\noindent \textbf{Vocoder.} 
Ultimately, the output features $\vec{\hat{x}}$ are fed into a vocoder backbone, specifically HiFi-GAN~\cite{kong2020hifigan} here, to reconstruct the features into the singing waveform $\vec{\hat{y}}$. 

\noindent \textbf{Loss Function.} 
For the loss function of the entire \mbox{SingOMD}, we adopt the same settings as those used in \cite{kong2020hifigan}, including GAN loss $L_{\text{Adv}}(G; D)$, feature matching loss $L_{\text{FM}}(G; D)$ and mel-spectrogtams loss $L_{\text{Mel}}$, where $G$ and $D$ represent the generator the discriminator respectively.
% $\begin{aligned} 
% \small
% &  
% \mathcal{L}_G=\mathcal{L}_{A d v}(G ; D)+\lambda_{f m} \mathcal{L}_{F M}(G ; D)+\lambda_{\text {mel }} \mathcal{L}_{M e l}(G) \\ 
% & 
% \mathcal{L}_D=\mathcal{L}_{\text {Adv }}(D ; G)
% \end{aligned}$

%\subsubsection{Extract discrete token}
\noindent \textbf{Discrete Representations.}
%After training SingAMD on the singing resynthesis task, to 
To obtain singing-adapted multi-resolution discrete representations, we apply K-means clustering individually on the continuous features $\hat{x}^{(t)}, \hat{x}^{(t-1)}, \ldots, \hat{x}^{(0)}$ from the resampling module corresponding to the desired resolutions for each $\hat{x}$ we want to obtain. This process results in obtaining corresponding discrete features for each desired resolution of $\hat{x}$.

\section{Experiments}

%When employing the SingOMD method for feature extraction, we initially train a SingOMD model to extract discrete representations from singing audio. 
To assess the effectiveness of discrete representations constructed via SingOMD, we conduct experiments on two tasks: singing sesynthesis and singing voice synthesis, and evaluate the synthesized audio accordingly.  

\subsection{SingOMD Training} \label{sec: vocoder-setting}
%\subsubsection{Dataset}
We first train SingOMD to construct discrete representations. 
%\noindent \textbf{Dataset:}
%The datasets utilized for training SingOM in singing resynthesis, that can be called as MixData, 
The training datasets comprise ACE-Opencpop~(130 hours)~\cite{shi2024singing}, OpenSinger~(50 hours)~\cite{huang2021multi}, M4Singer~(29.8 hours)~\cite{zhang2022m4singer}, and Opencpop~(5.2 hours)~\cite{wang22opencpop}, collectively amounting to approximately 210 hours. Notably, ACE-Opencpop, OpenSinger, and M4Singer are multi-singer datasets. 
We follow the predefined data splits for ACE-Opencpop and Opencpop. 
For M4Singer and OpenSinger, we %lack such divisions, prompting us to adopt a uniform approach for dataset segmentation across all datasets. We 
allocate the first 200 entries as the validation set, 201 and 250 entries as the test set respectively, and the remainder as the training set.

%\subsubsection{Experiment Setup}
%\noindent \textbf{Experiment Setup:}
%In the experiments training the SingOMD model, 
We choose HuBERT~\cite{hsu2021hubert}, one of the most prominent speech SSL models, as the pre-trained SSL model in our SingOMD. To better highlight the superiority of our approach, we opted for HuBERT base, featuring a 12-layer transformer with a hidden feature dimension of 768.
The resampling module consists of a transfer encoder and multiple Resampler blocks. The transfer encoder is comprised of a \texttt{Conv1d} with both input and output channels set to 512, a kernel size of 7, and a stride of 1. The parameters for each Resampler are identical to those of the sampling module described in \cite{shi2024multiresolution}. The kernel size and stride of \texttt{Conv1d} and \texttt{ConvTranspose1d} in Resampler are both 1. The residual coefficient $w_{\text{res}}$ in the resampling module is set to $\sqrt{0.4}$. 
Three resolutions are set at most, similar to the setting in ~\cite{shi2024multiresolution}.
In the vocoder, we follow the settings of HiFi-GAN~\cite{kong2020hifigan} and substitute spectrograms with features generated by resampling module.
To obtain discrete representations, the cluster number in K-means is set to 1024.
The training is performed using an NVIDIA 3090 GPU with a batch size of 16 for 250000 steps. We use the Adam optimizer with a learning rate of $2\times10^{-4}$. 
All experiments are conducted within the Parallel WaveGAN~\cite{Hayashi2021ESPnet2TTSET}.\footnote{\scriptsize{\url{https://github.com/kan-bayashi/ParallelWaveGAN}}}

\subsection{Evaluation Dataset and Metrics}
%\subsubsection{Dataset}
\noindent \textbf{Dataset.}
The Opencpop dataset serves as the benchmark for evaluating the quality of SingOMD tokens and other discrete representations. In our quality assessment experiments concerning discrete tokens, we follow the default segmentation of the Opencpop dataset. 

%\subsubsection{Metrics}
\noindent \textbf{Metrics.}
The quality of discrete tokens is assessed based on the quality of audio segments generated in both singing resynthesis and singing voice synthesis tasks using discrete tokens. We employ both subjective and objective metrics to evaluate the quality of these audio segments. 
The objective metrics include Mel cepstral distortion (MCD), logarithmic F0 root mean square error (F0 RMSE), semitone accuracy (S. Acc.), and Voice/Unvoice Error (V/UV E.), consistent with previous works~\cite{shi2022muskits, wu2023phoneix, wu2023systematic}. For subjective metrics, we utilize the Mean Opinion Score (MOS) approach, where 30 samples from each system are evaluated by 20 professional annotators on a 5-point scale, with 1 indicating an unreasonable synthesis and 5 signifying a synthesis indistinguishable from a real human voice.

\subsection{Evaluation setup on singing resynthesis task}
The singing resynthesis task directly converts the provided discrete tokens back into audio through a vocoder.

%\subsubsection{Baseline}
\noindent \textbf{Baselines.}
The compared baseline models, including both single stream tokens and multi-stream tokens configurations and vocoders as unit HiFiGAN~\cite{lee-etal-2022-direct, yan-etal-2023-espnet, hayashi2020discretalk}, are provided by the Interspeech2024 Challenge as follows: 
\begin{itemize}
    \item HuBERT-base/3: Single stream discrete tokens from the $3^{\text{rd}}$ layer of HuBERT base.
    \item HuBERT-base/sum: Single stream discrete tokens from weighted sum features of all layers in HuBERT base.
    \item HuBERT-base/3+10+11: Multi-stream discrete tokens from $3^{\text{rd}}$, $10^{\text{th}}$, $11^{\text{th}}$ layers of HuBERT base, top 3 weighted in weighted sum.
\end{itemize}

%\subsubsection{Experiment Setup}
\noindent \textbf{Experiment Setup.} 
In the singing resynthesis experiments, the vocoder employed is unit HiFi-GAN, which includes an additional embedding layer for input discrete tokens compared to HiFi-GAN. The rest of its architecture and parameters remain consistent with the official setting. The embedding layer is configured with 512 channels. In scenarios involving multiple stream tokens input, we utilize embedding layers to embed each stream separately and then integrate them through a weighted sum. The training settings align with Section~\ref{sec: vocoder-setting}.
% In training setting, unit HiFi-GAN is trained using a NVIDIA 3090 GPU with a batch size of 16 for 250000 steps. We use the Adam optimizer with a learning rate of $2\times10^{-4}$.
% All experiments are conducted within the Parallel WaveGAN.

\subsection{Experiment results on singing resynthesis}

\begin{table*}[h]
    \centering
    \caption{Comparison of discrete singing resynthesis on Opencpop. 95\% confidence intervals are reported in parentheses.}
    \label{tab:rs}
    \resizebox{0.9\linewidth}{!}{
    \begin{tabular}{l|l|c|c|cccccc} 
        \toprule
        & \textbf{Method} &\textbf{SSL} & \textbf{Resolution} & \textbf{MCD $\downarrow$} & \textbf{F0 RMSE $\downarrow$} & \textbf{S. ACC.$\uparrow$} & \textbf{VUV Error $\downarrow$} & \textbf{MOS $\uparrow$} &   \\
        \midrule
       1& Baseline & HuBERT-base/3 & (20) & 8.7103 & 0.2192 & 25.40\% & 9.93\% & 2.46 ($\pm$ 0.06) & \\
       2& Baseline & HuBERT-base/3+10+11 & (20) & 8.8802 & 0.2922 & 27.42\% & 8.74\% & 2.34 ($\pm$ 0.05) & \\
       3& Baseline & HuBERT-base/sum & (20) & 7.6427 & 0.1847  & 38.90\% & \textbf{7.66\%} & 2.78 ($\pm$ 0.06) & \\
        \midrule
        4& SingOMD (ours) & HuBERT-base/sum & (20,) & 6.9693 & 0.2167  & 60.32\% & 8.24\% & 3.39 ($\pm$ 0.06) & \\
        5& SingOMD (ours) & HuBERT-base/sum & (20, 40) & 6.6414  & \textbf{0.1806} & 64.02\% & 8.41\% & 3.48 ($\pm$ 0.06) & \\
        6& SingOMD (ours) & HuBERT-base/sum & (20, 40, 80) & \textbf{6.5766 } & 0.1828  &\textbf{ 64.83\%} & 8.16\% & \textbf{3.55 ($\pm$ 0.07)} & \\
        % MRHuBERT-base/6+12+17 & (20, 40, 80) & & & & \\
        \midrule
        7 & Ground Truth & - & - & - & - & - & - & 4.66 $\pm$ 0.06 & \\
        \bottomrule
    \end{tabular}
    }
\end{table*}
We conduct a comparison experiment on different discrete tokens and ablate on resolutions of SingOMD tokens.

\noindent \textbf{ \textit{Comparison of different discrete tokens.}} 
Table~\ref{tab:rs} shows the performance of different discrete tokens. "Resolution", refers to the resolution of the corresponding discrete representations. In experiments, our SingOMD tokens take only one stream for each resolution.
Comparing results from rows 3, and 4, it's evident that employing only a transfer encoder without introducing multi-resolution significantly enhances the synthesized outcomes. It also confirms the effectiveness of the weighted sum approach in rows 1 and 3. 
Furthermore, it's observed that both utilize 3-stream discrete tokens in rows 2 and 6. However, the quantity of tokens extracted from the SSL model in row 2 is nearly double that of row 6 due to different resolutions. Despite this difference, our approach still significantly outperforms in all metrics.
Additionally, a comparison between rows~1 and 2 illustrates that merely increasing the number of token streams does not lead to an improvement. In contrast, the inclusion of information from more resolutions results in substantial enhancements in rows 4-6.
These results demonstrate the superiority of SingOMD in extracting singing-oriented discrete tokens.

\noindent \textbf {\textit{Ablation on resolution.}} 
To investigate the impact of resolution on SingOMD, experiments were conducted using discrete tokens of varying resolutions. 
The results from rows 4 and 6 indicate that when transitioning from a single resolution to multiple resolutions, all metrics show significant improvement.
With the increase of resolutions in discrete tokens, as shown from rows~4-6, MCD, F0 RMSE, S. ACC., and MOS follow the improvement of resolutions or remain comparable results while VUV Error shows no apparent correlation.
It suggests that incorporating information from more resolutions can enhance the informational content of discrete features. 
However, comparing rows~4-5 and rows~5-6, it seems that further increasing resolutions does not yield as pronounced effects as the initial shift.
This finding suggests a trade-off between the quality of synthetic vocals and the quantity of tokens which represents the computational efficiency.

\subsection{Evaluation setup on singing voice synthesis}
We also evaluate the effectiveness of our singing-oriented discrete tokens on the SVS task. Specifically, We first train a vocoder using discrete tokens as input. Then, we train a discrete token-based acoustic model to directly predict discrete tokens from musical scores. Finally, the acoustic model and vocoder are cascaded as an SVS system.

%\subsubsection{Baseline}
\noindent \textbf{Baselines.}
For all systems utilizing discrete tokens, the acoustic model employed for predicting discrete features is an RNN-based model~\cite{shi2021sequence} provided by the Interspeech2024 Challenge. 
To further explore the effectiveness of our method, we do not use any additional information in discrete tokens systems. 
For systems using Mel-spectrograms, we utilize XiaoiceSing~\cite{lu2020xiaoicesing}, a classic transformer based model, as the acoustic model. 
All acoustic models above are available in ESPnet-Muskits~\cite{shi2022muskits}.\footnote{\scriptsize{\url{https://github.com/espnet/espnet}}} 
All vocoders are either HiFi-GAN or unit HiFi-GAN, consistent with previous experiments.

The SVS baselines are as follows:
\begin{itemize}
    \item Mel-septrograms: XiaoiceSing as acoustic model, HiFi-GAN as vocoder and mel-spetrograms as intermediate features.
%    \item Challenge baseline~(f0): RNN with duration predictor and pitch predictor, single stream discrete tokens from the $6^{th}$ layer of WavLM large~\cite{chen2022wavlm}.
    \item HuBERT-base/3: RNN with duration predictor, single stream from the $3^{rd}$ layer of HuBERT base.
    \item HuBERT-base/3+10+11: RNN with duration predictor, multi stream discrete tokens from the $3^{\text{rd}}$, $10^{\text{th}}$ , and $11^{\text{th}}$ layers of HuBERT base.
\end{itemize}

\noindent \textbf{Experiment Setup.}
In SVS experiments, all acoustic models including RNN and XiaoiceSing and the training parameters, adhere to the suggested settings specified in ESPnet Opencpop recipe.\footnote{\scriptsize{\url{https://github.com/espnet/espnet/tree/master/egs2/opencpop/svs1}}} 
For SingOMD systems, SingOMD tokens are chosen in resolution [20, 40, 80]. All acoustic models are trained using the Adam optimizer with a learning rate of $1\times10^{-3}$ on a NVIDIA 3090 GPU with a batch size of 16 for 350 epochs. We choose the best model from the validation set.
The configurations for vocoders are consistent with those detailed in \ref{sec: vocoder-setting}, following the official HiFi-GAN settings. 
All experiments are conducted within the ESPnet framework.

\subsection{Experiment results on singing voice synthesis}
The results in Table \ref{tab:svs} demonstrate that \mbox{SingOMD} achieves the best grades among all systems using discrete tokens in all metrics. 
%Notably, our method surpasses the challenge baseline even when utilizing the less effective HuBERT base model and without the use of a pitch predictor. This strongly validates the superiority of the discrete tokens extracted by SingOMD. 
Furthermore, our approach achieves comparable MOS to the baseline system with Mel spectrograms and improves notably in F0 RMSE which validate our hypothesis that our model requires long-duration information in pitch. And the deterioration of MCD in SingOMD is reasonable, due to information loss during discretization. 
Although there is an undeniable gap between discrete systems and Ground-Truth, the results still underscore the efficacy and potential of \mbox{SingOMD} in enhancing the quality and effectiveness of singing voice synthesis.

\begin{table}[t!]
    \centering
    \caption{Comparison of SVS performance on Opencpop. 95\% confidence intervals are reported in parentheses.}
    \label{tab:svs}
    \centering
    \resizebox{\linewidth}{!}{
    \begin{tabular}{l|cccc} 
        \toprule
        \textbf{Model}  & \textbf{MCD $\downarrow$} & \textbf{F0 RMSE $\downarrow$} & \textbf{MOS $\uparrow$} &   \\
        \midrule
        Mel spectrogram & \textbf{6.9283 } & 0.2610  & 3.04 $\pm$ 0.06 & \\
        \midrule
%        Challenge baseline & 8.4199  & \textbf{0.2249 } & & \\
        HuBERT-base/3 & 9.5528  & 0.2321  & 2.34 $\pm$ 0.06 & \\
        HuBERT-base/3+10+11 & 9.7585  & 0.3200  & 2.34 $\pm$ 0.05 & \\
        \midrule
        DiscreteSVS+SingOMD & 7.7234  &\textbf{ 0.1941}  & \textbf{3.10 $\pm$ 0.06} & \\
        \midrule
        Ground Truth & - & - & 4.66 $\pm$ 0.06 & \\
        \bottomrule
    \end{tabular}
    }
\end{table}

\section{Conclusion}

This paper proposes SingOMD, a novel method to construct singing-oriented discrete representations for singing generation by leveraging speech SSL models. SingOMD first alleviates domain gaps between speech and singing by adapting the continuous features from hidden layers of speech SSL for singing through a singing audio resynthesis process. Moreover, a resampling module is incorporated to capture multi-resolution richer features. These adapted multi-resolution features are then discretized via K-means to form our singing-oriented discrete representations. Extensive experiments demonstrate the robustness of these representations in singing vocoders. They also enhance the efficiency and effectiveness of singing voice synthesis when integrated with a discrete singing acoustic model.
% \section{Acknowledgements}

% Experiments of this work used the Bridges2 system at PSC and Delta system at NCSA through allocations CIS210014 and IRI120008P from the Advanced Cyberinfrastructure Coordination Ecosystem: Services \& Support (ACCESS) program, supported by NSF grants \#2138259, \#2138286, \#2138307, \#2137603, and \#2138296. We would like to thank Shengyuan Xu and Pengcheng Zhu for their support in the data license.

% \section{Acknowledgements}

% \ifinterspeechfinal
%      The INTERSPEECH 2023 organisers
% \else
%      The authors
% \fi
% would like to thank ISCA and the organising committees of past INTERSPEECH conferences for their help and for kindly providing the previous version of this template.

% \bibliographystyle{IEEEtran}
% \bibliography{mybib}

\section{Acknowledgements}
This work was partially supported by the  National Natural Science Foundation of China (No. 62072462) and the Beijing Natural Science Foundation (No. L233008).

\section{References}
{
\printbibliography
}

\end{document}